# SCIENTIFIC REPORTS

**OPEN**

# Generalized phase mixing: Turbulence-like behaviour from unidirectionally propagating MHD waves

Norbert Magyar, Tom Van Doorsselaere & Marcel Goossens

We present the results of three-dimensional (3D) ideal magnetohydrodynamics (MHD) simulations on the dynamics of a perpendicularly inhomogeneous plasma disturbed by propagating Alfvénic waves. Simpler versions of this scenario have been extensively studied as the phenomenon of phase mixing. We show that, by generalizing the textbook version of phase mixing, interesting phenomena are obtained, such as turbulence-like behavior and complex current-sheet structure, a novelty in longitudinally homogeneous plasma excited by unidirectionally propagating waves. This study is in the setting of a coronal hole. However, it constitutes an important finding for turbulence-related phenomena in astrophysics in general, relaxing the conditions that have to be fulfilled in order to generate turbulent behavior.

Phase mixing of Alfvén waves, in the context of astrophysics, was originally suggested as a candidate mechanism to heat both the open magnetic field corona (e.g. coronal holes) and coronal loops[1]. Heating is achieved through small scale generation: Alfvén waves excited in the lower solar atmosphere propagate outward, and neighbouring oscillating field lines get out of phase if a varying Alfvén speed profile is present transverse to the propagation direction. Once the resulting scales are small enough, viscous and/or resistive dissipation occurs. Another, increasingly popular wave heating (AC) mechanism (it can also be viewed as a DC mechanism in case of quasi-static driving[2]) proposed for heating the solar corona[3] is MHD turbulence[4–8], which has general applications in astrophysics[9,10], as well as in the physics of fusion and numerous industrial applications[11,12]. The underlying idea for energy conversion is the same as in phase mixing: small scales are generated by the nonlinear cascade of wave energy, until the dissipation range is reached. For turbulence to be sustained in incompressible MHD, counterpropagating waves need to be present in the medium[13]. One often uses the Elsässer variables ($\mathbf{z}^{\pm} = \mathbf{v} \pm \mathbf{b}/\sqrt{4\pi\rho}$, where $\mathbf{v}$ is velocity, $\mathbf{b}$ is magnetic field and $\rho$ is plasma density) to better differentiate between outward (−) and inward (+) propagating Alfvén waves, for an outwardly directed mean magnetic field $\mathbf{B}_0$[9,14,15]. Then the necessity for counterpropagating wave presence can be easily seen from the Elsässer formulation of the (incompressible) MHD equations:[16]

$$\nabla \cdot \mathbf{z}^{\pm} = 0,$$
$$\frac{\partial \mathbf{z}^{\pm}}{\partial t} + (\mathbf{z}^{\mp} \cdot \nabla)\mathbf{z}^{\pm} = -\frac{1}{8}\nabla(\mathbf{z}^{+} - \mathbf{z}^{-})^2, \quad (1)$$

here the essential nonlinearity comes from the advective derivatives, requiring both $\mathbf{z}^{\pm}$ to be nonzero, or in other words, counterpropagating Alfvén waves. Note that the previously described phenomenology, also known as the Alfvén effect, is strictly valid only in the framework of incompressible MHD. In plasma with a finite speed of sound, compressible modes, i.e. fast and slow MHD waves are present. However, in a plasma with an inhomogeneous background or equilibrium, the MHD waves are linearly coupled, or more descriptively, they have mixed properties[17,18]. This also implies that it is not possible to decompose[19] the perturbations into Alfvén, fast, and slow waves, as there are no pure MHD wave modes. It was shown recently that the Alfvén effect is no longer a valid

Centre for mathematical Plasma Astrophysics (CmPA), KU Leuven, Celestijnenlaan 200B bus 2400, 3001, Leuven, Belgium. Correspondence and requests for materials should be addressed to N.M. (email: norbert.magyar@kuleuven.be)

phenomenology for isothermal compressible MHD turbulence[20], but needs generalization. This implies that the generation of turbulence is no longer restricted to counterpropagating Alfvén waves in compressible MHD. In the compressible case, by using compressible Elsässer variables (defined as above but with varying density) we can still keep the Elsässer formalism, leading to different equations[16] than in Eqs. 1:

$$\frac{\partial \ln \rho}{\partial t} + (\mathbf{z}^\pm \cdot \nabla)\ln \rho = -\frac{1}{2}\nabla(3\mathbf{z}^\pm - \mathbf{z}^\mp), \frac{\partial \mathbf{z}^\pm}{\partial t} + (\mathbf{z}^\mp \cdot \nabla)\mathbf{z}^\pm$$
$$= \pm\frac{1}{4}(\mathbf{z}^+ - \mathbf{z}^-)\left(\frac{\partial \ln \rho}{\partial t} + (\mathbf{z}^\pm \cdot \nabla)\ln \rho\right)$$
$$-\frac{1}{8}\nabla(\mathbf{z}^+ - \mathbf{z}^-)^2 - \left[c_s^2 + \frac{1}{8}(\mathbf{z}^+ - \mathbf{z}^-)^2\right]\nabla \ln \rho, \quad (2)$$

where $c_s^2 = \partial p / \partial \rho$ is the speed of sound. For the sake of simplicity in this introductory section, the equations above were derived using a polytropic equation of state $p = p_0(\rho/\rho_0)^\gamma$, where $\gamma$ is the polytropic index. Note that by letting density variations vanish in Eqs. 2, we readily recover the incompressible case of Eqs. 1. Also note that, by considering a weakly compressible case, nonlinearity arises essentially from the same nonlinear advective terms as in the incompressible case, involving both $\mathbf{z}^\pm$. The crucial difference is, however, that generally we can no longer interpret the Elsässer variables as representing strictly parallel and anti-parallel propagating Alfvén waves, due to contributions from the compressible modes, or traduced in the fully coupled (inhomogeneous) case, due to the mixed compressible/Alfvén properties of the waves. In the incompressible case, starting from a spectrum of purely outgoing Alfvén waves (e.g. as in a coronal hole), a source of incoming waves is necessary for nonlinear interactions to sustain MHD turbulence. This source can be the inhomogeneities, so-called large-scale gradients, parallel to the mean magnetic field in the plasma (e.g. gravitational stratification in a coronal hole), causing reflections of the waves[2,21–23]. The parametric decay instability, a nonlinear instability of Alfvén waves, is also capable of generating backward propagating Alfvén waves, being more effective in the high-amplitude and/or high-frequency regime[24,25]. However, if we consider compressible modes, as stated previously, nonlinear interactions ($\mathbf{z}^\pm \neq 0$) are no longer restricted to counterpropagating Alfvén waves: in an inhomogeneous medium, propagating MHD waves have mixed properties, and appear differently (predominantly Alfvén or fast characteristics) in different regions of the plasma, dictated by the local inhomogeneity[18,26]. This can be viewed, in the Elsässer formalism, as a unidirectionally propagating wave presenting both $\mathbf{z}^+$ and $\mathbf{z}^-$ fields to a varying degree, depending on the local inhomogeneity. That is, a wave with mixed properties is necessarily described using both $\mathbf{z}^\pm$. The $\mathbf{z}^+$ and $\mathbf{z}^-$ fields of waves with mixed properties no longer represent Alfvén waves propagating in opposite directions: they propagate in the same direction and with the same phase speed: they describe a unidirectionally propagating wave. In this sense, nonlinear terms represent self-deformation of waves. In this paper we aim to verify these claims and study a perpendicularly inhomogeneous medium perturbed by unidirectionally propagating Alfvénic waves, by employing numerical simulations. Here we would like to point out the meaning of 'Alfvénic', as formulated by:[26,27] it describes waves which have largely Alfvén characteristics, however, due to plasma inhomogeneity they are not pure Alfvén waves, as compression is also present. Alfvénic waves are an example of MHD waves with mixed properties.

In the following, we describe the numerical setup used in our simulations under the Methods section, followed by the Results. The main conclusions are drawn in the Conclusions section.

## Methods

We employ 3D ideal MHD numerical simulations to test the previously described scenario of unidirectionally propagating Alfvénic waves in a perpendicularly inhomogeneous medium, using MPI-AMRVAC[28,29], which solves the fully nonlinear, ideal MHD equations in 3D Cartesian geometry:

$$\frac{\partial \rho}{\partial t} + \nabla \cdot (\mathbf{v}\rho) = 0,$$
$$\frac{\partial (\rho\mathbf{v})}{\partial t} + \nabla \cdot (\mathbf{v}\rho\mathbf{v} - \mathbf{BB}) + \nabla p_{tot} = 0,$$
$$\frac{\partial E}{\partial t} + \nabla \cdot (\mathbf{v}E - \mathbf{BB} \cdot \mathbf{v} + \mathbf{v}p_{tot}) = 0,$$
$$\frac{\partial \mathbf{B}}{\partial t} + \nabla \cdot (\mathbf{vB} - \mathbf{Bv}) = 0 \quad (3)$$

where $p_{tot} = p + \mathbf{B}^2/2$ is the total pressure, and $E = \frac{p}{\gamma - 1} + \rho\frac{\mathbf{v}^2}{2} + \frac{\mathbf{B}^2}{2}$ is the total energy density. We set the adiabatic index, $\gamma$ to $\frac{5}{3}$. We supplement the MHD equations with the ideal gas law, $p = \rho\frac{k_B}{\mu m_H}T$, where the average mass per particle (in units of hydrogen atom mass $m_H$) is $\mu = 0.6$ for coronal abundances. Here, the magnetic field is measured in units for which the magnetic permeability is 1. Note that, unlike in the introductory section, a full energy equation is used, and the equations are solved for the usual variables ($\rho$, $p$, $\mathbf{v}$, $\mathbf{B}$), from which the Elsässer variables are calculated for analysis. The finite volume method uses the TVD second-order accurate solver and Woodward slope limiter. The solenoidal constraint on the magnetic field ($\nabla \cdot \mathbf{B} = 0$) is enforced using Powell's scheme. The numerical domain (see Fig. 1), aimed to represent a thin elongated section of a coronal hole is $1 \times 1 \times 20$ Mm in size, discretized uniformly with $512 \times 512 \times 128$ numerical cells. This translates in cell sizes in the $x$–$y$ plane of ≈2 km, putting heavy constraints on the CFL timestep, and limiting the achievable resolution to its present value. We consider a straight, homogeneous magnetic field of $B_0 = 5$ G directed along the $z$-axis. Gravity is neglected, thus there is no stratification along the magnetic field. In this way, we eliminate possible reflections of unidirectionally propagating Alfvénic waves on equilibrium gradients along the magnetic field. Besides, having in mind the typical density scale height in a 1 MK corona (50 Mm), it is a good approximation for the present study. The plasma $\beta = \frac{2p}{B_0^2} \approx 0.15$ is constant throughout the domain at $t = 0$. The large-scale gradient transverse to the magnetic field is given by inserting, on a uniform background density of $\rho_0 = 2 \cdot 10^{-13}$ kg m$^{-3}$ random (in amplitude, position, and width) Gaussian density enhancements in the $x$–$y$ plane, uniform along the $z$-axis:

$$\rho(x, y, z) = \rho_0 + \sum_{i=1}^{250} A_i \exp^{-\left[(x-x_i)^2 + (y-y_i)^2\right]/\sigma_i}, \tag{4}$$

where $A_i \to [0, 5\rho_0]$, $x_i \to [-0.5 \text{ Mm}, 0.5 \text{ Mm}]$, $y_i \to [-0.5 \text{ Mm}, 0.5 \text{ Mm}]$, $\sigma_i \to [0, 5 \cdot 10^{-3} \text{ Mm}^2]$ are amplitude, position, and Gaussian width, respectively, randomly chosen (uniform distribution) within their respective limits. These density enhancements aim to represent random perpendicular density variations in a coronal hole, in equilibrium. These kind of inhomogeneities, in reality, might arise for example from small localized heating events, leading to chromospheric evaporation. The structure is periodic in the $x$–$y$ plane, as we set periodic boundaries laterally. Periodicity is ensured by letting the contribution of each Gaussian density enhancement 'pass' through the closest $x$-axis and $y$-axis periodic boundaries. The resulting mean density is ≈$1.2 \cdot 10^{-12}$ kgm$^{-3}$, while the peak value is ≈$3 \cdot 10^{-12}$ kgm$^{-3}$. At the top boundary, we use a Neumann-type zero-gradient 'open' boundary condition for all variables. This is essential, as we want to exclude or minimize as much as possible the generation of reflected, counterpropagating waves. Tests with homogeneous density runs show maximum 0.5% reflection of the incident wave energy on the top boundary. At the bottom, we employ a wave driver with properties intending to mimic observed Alfvénic waves in coronal holes[23,30]. In this sense, we use a superposition of 10 sinusoidal waves:

$$\mathbf{v}_x(x, y, 0, t) = \sum_{i=1}^{10} U_i \sin(\omega_i t),$$
$$\mathbf{v}_y(x, y, 0, t) = \sum_{i=1}^{10} V_i \sin(\omega_i t), \tag{5}$$

each with a definite angular frequency $\omega_i$, velocity amplitudes $U_i$ and $V_i$, obtained randomly from the observed log-normal distributions[23], and random direction in the $x-y$ plane, resulting from the different random velocity amplitudes for $x$ and $y$-axis components, leading to transverse, Alfvénic waves propagating upwards on the $z$-axis. Note that the driver is time but not space-dependent, i.e. the whole boundary is driven equally. The resulting rms (root mean-square) velocity amplitude is ≈12 kms$^{-1}$. Initially, in the simulation domain, the average sound speed $c_s = \sqrt{k_B T / \mu m_H}$ is ≈ 120 kms$^{-1}$, while the average Alfvén speed $V_A = B_0 / \sqrt{\rho}$ is 440 kms$^{-1}$. These values show us that we are in the weakly compressible, subsonic regime of flow ($\mathbf{v}_{rms} < \langle c_s \rangle < \langle V_A \rangle$).

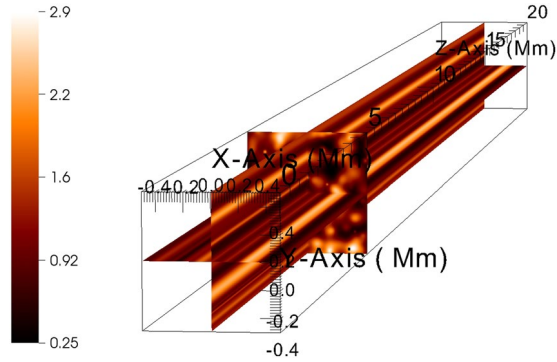

**Figure 1.** Three-slice plot of the 3D numerical domain. Note that the inhomogeneity is in the $x-y$ plane. The domain is uniform along the $z$-axis. The density color legend is shown in units of $10^{-12}$ kg m$^{-3}$.

### Results

We run the simulation for 1000 s. During this time, perturbations originating from the boundary driver propagate upwards (on the $z$-axis) and interact with the inhomogeneities. The initial (equilibrium) transverse structure is quickly destroyed by the propagating Alfvénic waves, and transformed into a cross-section presenting structures on a large range of scales, reminiscent of turbulence (see Fig. 2). This observation was at the basis of a previous study which doubts the existence of packed thin individual and dynamically independent magnetic elements in the solar corona, so-called 'coronal strands' or filaments, when perturbed by propagating Alfvénic waves[31]. The current density shows the same turbulent structure in the cross-section with numerous current sheets forming (see Fig. 3) and getting dissipated by numerical resistivity. Note how the current in individual strong current sheets is more than an order of magnitude stronger than the average current density. The peak values in the current density depend on the resolution and numerical resistivity (intrinsic to the numerical scheme, not set), but should ultimately saturate at sufficiently high resolutions. We estimate the Lundquist number in the simulation to be of the order $10^5$–$10^6$. In test runs with half resolution, the peak values in the current density drop approximately to half as compared to the present setup, indicating that we did not reach the saturated state. The dissipation (heating) leads to an increase in the internal energy of the plasma. However, this represents only a $\approx$2.2% rise with respect to the equilibrium value of the internal energy at the end of the simulation time, being unable to change the plasma temperature significantly. The available energy flux from the driver is, on average, 50 Wm$^{-2}$, in agreement with the estimated energy content of the observed transverse waves in coronal holes[23]. However, only a small part of this energy gets deposited in the domain, as waves propagate through the open top boundary. On average, the energy flux decreases $\approx$1.5% from its original value as it leaves the domain. This results in a heating rate of $\approx$3.1 × 10$^{-7}$ erg cm$^{-3}$ s$^{-1}$, way below the $\approx$10$^{-4}$ erg cm$^{-3}$ s$^{-1}$ that would be required to balance radiative losses[32]. As we do not focus on the energetics (processes mainly parallel to the main magnetic field) but rather on the turbulent behavior (perpendicular to the main magnetic field) in this study, heat sources or sinks, thermal conduction or thin radiative losses are not included in the energy equation. The time and space-averaged transverse magnetic field fluctuation magnitude is $\approx$3.0% of B$_0$.

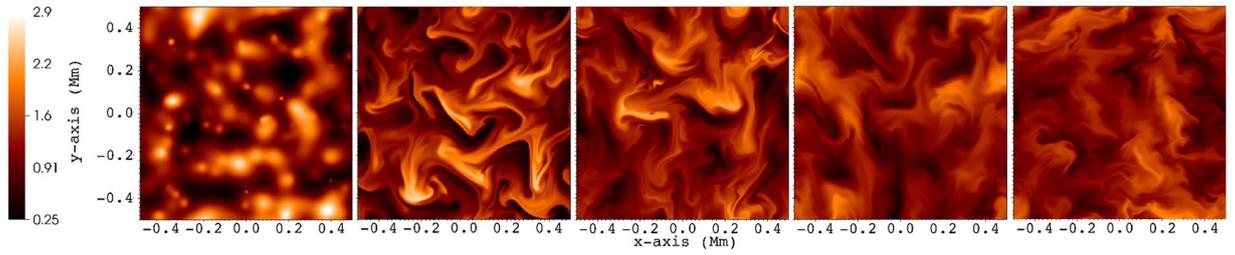

**Figure 2.** Evolution of density in a 2D slice taken at $z = 18$ Mm, shown in steps of 250 s (to the right), the first plot showing the equilibrium transverse structure ($t = 0$ s). The density color legend is shown in units of $10^{-12}$ kg m$^{-3}$.

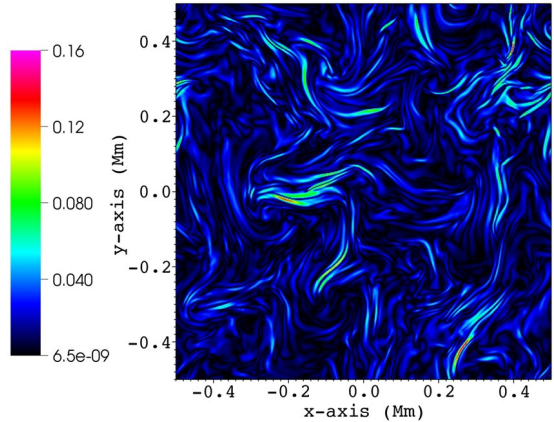

**Figure 3.** Absolute value of current density (user units) along the $z$-axis ($j_z$) in a 2D slice taken at $z = 18$ Mm and $t = 500$ s.

We calculated the 1D power spectra of magnetic and kinetic energy, density, pressure (shown in Fig. 4), and $z^\pm$ (not shown, for clarity) in a plane perpendicular to the mean magnetic field B$_0$. Note that the small scale generation or cascade is purely perpendicular to the mean magnetic field B$_0$. We use the fft routine of IDL to obtain $\left|f_{k_x,k_y}\right|^2$, where f = **v, b,** $z^\pm$, $\rho$, $p$. The magnetic and kinetic energy spectrum is then

$$E_{KM}(k_\perp) = \sum_{k_x,k_y}\left(\left|\mathbf{v}_{k_x,k_y}\right|^2 + \left|\mathbf{b}_{k_x,k_y}\right|^2\right), \quad k_\perp = \sqrt{k_x^2 + k_y^2} \tag{6}$$

while the calculation is similar for the other variables. Note, however, that $E_{KM}$ is not the total energy: in compressible MHD the total energy contains also the internal energy[16]. The time-averaged $E_{KM}$ spectrum ($k_\perp^{-2.54}$) is steeper than all well-known MHD inertial range spectra (e.g. $k_\perp^{-5/3}$ or $k_\perp^{-2}$ in strong or weak incompressible MHD turbulence, respectively). Lower resolution test simulation runs show similar spectral slopes, however, the inertial range is shorter, due to enhanced numerical dissipation. However, the present inertial range length (less than a decade in wavenumber) and the high variability of its slope in time suggests that the slope values should be treated carefully. Still, the obtained inertial range slopes suggest new dynamics. Indeed, our dynamics resembles an unsteady, imbalanced weak MHD turbulence, with tempo-spatial varying energy ratio of the $z^\pm$ fields. We use the term 'weak' based on the observation that the cascade is essentially perpendicular to the background $B_0$ field. Currently, there are no analytical or even phenomenological models describing this scenario in compressible MHD. No attempt is made here to develop such a model. The closest incompressible MHD turbulence description to the present scenario is that of imbalanced weak MHD turbulence[33,34]. Note, however, that there are still fundamental differences with respect to the latter: The nonlinear cascade in our system is not driven by counterpropagating Alfvén waves, but by the mixed properties of unidirectionally propagating MHD waves in an inhomogeneous plasma, presenting self-deformation. The inertial range spectra changes in time due to the varying boundary driver, and the varying presence of $z^+$ (see Fig. 5), thus the aforementioned varying $E^-/E^+$ ($E^\pm = |z^\pm(k_\perp)|^2$) Elsässer energy ratio. In Fig. 4 we also show the energy spectrum at $t = 500$s, which presents an inertial range slope close to that in weak MHD turbulence[9,35], $k^{-2}$. In imbalanced weak MHD turbulence[36], $E^\pm \propto k_\perp^{-2\pm\alpha}$, where $\alpha \neq 0$ and corresponds to different inertial range slopes for Elsässer energies. We get $\approx -1.85$ and $-2.01$ for the $E^+$ and $E^-$ inertial range slopes, respectively, at $t = 500s$. The time-average slopes are $\approx -2.08$ and $-2.8$, respectively. The pinning effect (converging $E^\pm$ spectra at the dissipation range, see[33]) of the Elsässer energies is observed, around $k_\perp = 0.02$ km$^{-1}$. There is a striking difference in the inertial range slopes of pressure and density: for a polytropic law, dimensional analysis leads us to similar $k^{-7/3}$ slopes for both variables, which should be valid even in weakly compressible cases, as the present one[9]. The deviation from this law comes from the use of a full energy equation instead of a polytopic relation, which allows us to set up a moderately inhomogeneous density profile while setting the pressure constant initially. In Fig. 5 the $z^+$ field is shown, as well as the $z^-$ contours. Note that these two fields do not represent two different, counterpropagating waves: they are the manifestations of MHD waves with mixed properties. The dominant component of the waves is the $z^-$ field, showing that the driven wave has a largely Alfvénic character[37]. The time and space-averaged amplitude of the $|\mathbf{z}^+|$ field is ≈14% of $|\mathbf{z}^-|$, hence the imbalanced nature of the turbulence. We can see that the $\mathbf{z}^\pm$ propagation directions coincide, and there appears to be no significant reflection on the top boundary (at $\mathbf{z} = 20$ Mm). Performing a $k-\omega$ diagram analysis shows a weak presence of reflected waves in $|\mathbf{z}^+|$, which originate either from the top boundary or get reflected on density perturbations inside the domain, as a second-order effect. However, these are negligible when compared to the non-reflected, Alfvénic $|\mathbf{z}^+|$ component. Also, $|\mathbf{z}^+|$ displays a bursty, patchy evolution, as expected from the high degree of perpendicular density inhomogeneity in the domain. This might explain the observed rapid variation of the calculated spectra. To further strengthen our point about the origin of the $|\mathbf{z}^+|$ field, we ran additional simulations, identical to the presented one, except removing the inhomogeneities in density for $\mathbf{z} \geq 10$ Mm, i.e. a sudden transition to the constant background density. In Fig. 6, the resulting $z^\pm$ fields are shown at $t = 30$ s, i.e. just before the first wavefront reaches the top $z$ boundary. We chose this early snapshot in order to partially exclude the waves reflected on the interface at $\mathbf{z} = 10$ Mm, which otherwise would dominate the $|\mathbf{z}^+|$ field and hide the $|\mathbf{z}^+|$ contribution of the outward propagating Alfvénic waves. The waves reflected on the interface are counterpropagating[2], and can be seen in Fig. 6 as the strong wavefronts in $|\mathbf{z}^+|$, originating from $\mathbf{z} = 10$ Mm. Note how the $|\mathbf{z}^+|$ component is practically absent in the homogeneous region $\mathbf{z} \geq 10$ Mm, but there's an even stronger $|\mathbf{z}^-|$, as a consequence of conservation of energy. In this sense, the $|\mathbf{z}^+|$ in the inhomogeneous region (excluding the contribution from the reflected counterpropagating waves at the interface) is a manifestation of a unidirectionally propagating MHD wave with mixed properties in a perpendicularly inhomogeneous medium and cannot be associated with a different wave.

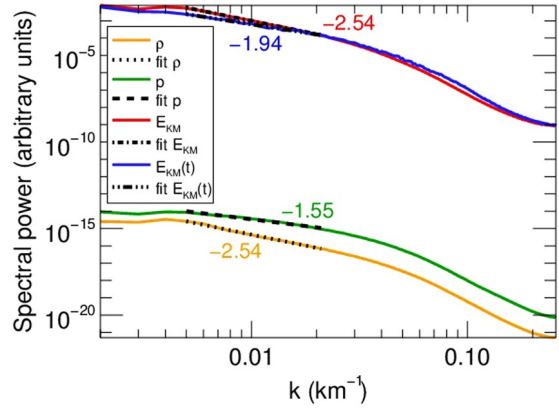

**Figure 4.** Spectra (power spectral density) of energy ('$E_{KM}$' - time averaged over 200−1000s; '$E_{KM}(t)$' - energy spectrum at $t = 500$ s), density, and pressure ('$\rho$, p' - time averaged over 200−1000s) calculated in a perpendicular slice at $z = 18$ Mm. Linear fits to the inertial range are also shown for each spectrum. Values for the slopes are also shown, in matching colors.

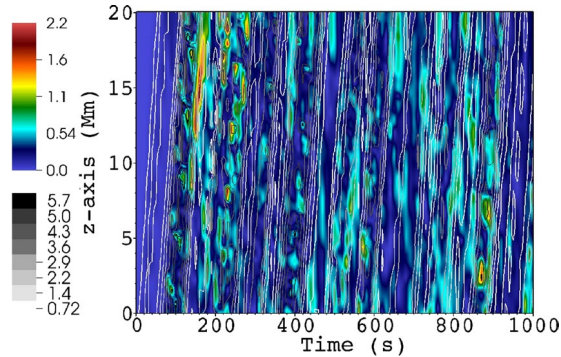

**Figure 5.** Time-distance map of $|z^-|$ (greyscale contour plot) and $|z^+|$ (color plot) along the segment $(x, y, z) = (0, 0, z)$. The step-like appearance of some wave-fronts is a visualization artifact, due to limited number of snapshots in time (100). Values are in user units.

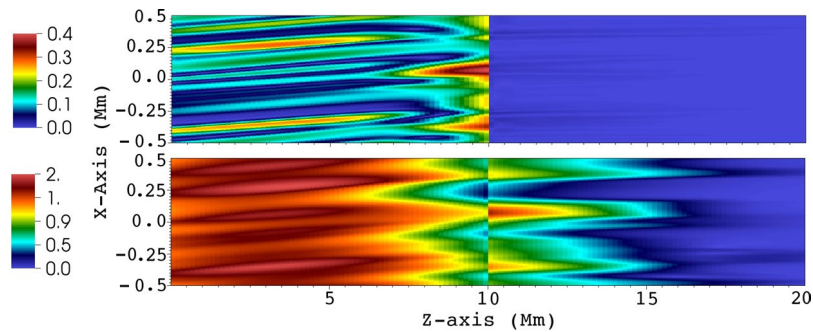

**Figure 6.** : Snapshots of $|\mathbf{z}^+|$ (top) and $|\mathbf{z}^-|$ (bottom) in the $(x, 0, z)$ plane at $t = 30$ s, of the setup with homogeneous background density after $z \geq 10$ Mm, with corresponding color legends on the sides. Values are in user units.

## Conclusion

We performed a first, full 3D MHD simulation of a generalized phase mixing scenario, with the varying Alfvén speed in the domain given by a random density profile perpendicular to the uniform and straight equilibrium magnetic field. The equilibrium is perturbed by upward-travelling Alfvénic waves (along the magnetic field), which originate from a driver at the lower $z$-axis boundary, modeled after the observed properties of waves in coronal holes. The waves leave the domain at the top $z$-axis boundary, and as we consider an initially uniform plasma along the magnetic field, no significant wave reflections occur. Hence we study unidirectionally propagating Alfvénic waves in a perpendicularly inhomogeneous plasma. The resulting dynamics are complex and turbulent in the cross-section, however the solution remains smooth parallel to the $z$-axis. Current sheets form in the process, displaying ribbon-like appearance along the main magnetic field, when viewed in 3D. These current sheets are dissipated due to numerical resistivity, and this leads to a small rise in the internal energy. This heating (at least in this setup) does not cause significant changes to the average temperature in the domain. The driven, propagating Alfvénic wave has mixed properties due to the inhomogeneities, which manifests as simultaneously nonzero $\mathbf{z}^\pm$ fields, presenting nonlinear self-deformation, in this sense. This nature of the MHD waves with mixed properties is at the origin of the complex, turbulent behaviour observed in the cross section, allowing for nonlinear cascade of wave energy to smaller scales of unidirectionally propagating waves. Let us emphasize the importance of the last sentence, as it is the main conclusion: nonlinear interactions, at the core of turbulent behaviour, are no longer restricted to counterpropagating waves in the scenario described in this paper. This study constitutes a first numerical demonstration of the recently realized[20] fact that the Alfvén effect is no longer valid in its current form in compressible MHD, but needs generalization. Thus, we simulated a potentially new type of MHD turbulence. The perpendicular spectrum of energy displays a power-law like appearance. This spectrum is highly variable in time, and its average slope for the inertial range is not in accordance with any presently available theory, being steeper than the well-known power laws of MHD turbulence. However, there is a clear separation of inertial and dissipation ranges, and the pinning effect, observed in numerical simulations of imbalanced weak MHD turbulence, is present. These findings might relax the criteria for the existence of turbulence-like cascade in plasma with a mean magnetic field and large-scale inhomogeneities perpendicularly to it. As this scenario is likely frequently present in astrophysical context, it might make turbulent behavior nearly ubiquitous.

## Acknowledgements

N.M. thanks Francesco Pucci and Vyacheslav Olshevskyi for fruitful discussions. N.M. acknowledges the Fund for Scientific Research-Flanders (FWO-VLaanderen). T.V.D. was supported by an Odysseus grant and received funding from the European Research Council (ERC) under the European Union's Horizon 2020 research and innovationprogramme (grant agreement No 724326). T.V.D. and M.G. acknowledges support by the Belspo IAP P7/08 CHARM network and the GOA-2015-014 (KU Leuven).

### Author Contributions
N.M. performed the simulations, analysis, and wrote the manuscript. N.M., T.V.D., and M.G. interpreted the results. All authors reviewed the manuscript.

### Additional Information
**Competing Interests:** The authors declare that they have no competing interests.

**Publisher's note:** Springer Nature remains neutral with regard to jurisdictional claims in published maps and institutional affiliations.